\documentclass{sigchi-ext}
\usepackage[T1]{fontenc}
\usepackage{textcomp}
\usepackage[scaled=.92]{helvet} 
\usepackage{graphicx} 
\usepackage{balance}  
\usepackage{booktabs} 
\usepackage{ccicons}  
\usepackage{ragged2e} 
\usepackage{subfig}

\def\plaintitle{Seniors' Media Preference for Receiving Internet Security Information: A Pilot Study} 
\def\emptyauthor{}

\title{Seniors' Media Preference for Receiving Internet Security Information: A Pilot Study}

\numberofauthors{6}
\author{%
  \alignauthor{%
    \textbf{Yousra Javed}\\
    \affaddr{Illinois State University} \\
    \email{yjaved@ilstu.edu} }\alignauthor{%
    \textbf{Boyd Davis}\\
    \affaddr{University of North Carolina Charlotte}\\
    \email{bdavis@uncc.edu} } \alignauthor{%
    \textbf{Mohamed Shehab}\\
    \affaddr{University of North Carolina Charlotte}\\
    \email{mshehab@uncc.edu} }
    }

\definecolor{linkColor}{RGB}{6,125,233}
\hypersetup{%
  pdftitle={\plaintitle},
  pdfauthor={\emptyauthor},
  bookmarksnumbered,
  pdfstartview={FitH},
  colorlinks,
  citecolor=black,
  filecolor=black,
  linkcolor=black,
  urlcolor=linkColor,
  breaklinks=true,
}

\begin{document}


\maketitle

\RaggedRight{} 

\begin{abstract}
Due to the increasing use of Internet by older adults and their low computer and Internet security literacy, their susceptibility to online fraud has also increased. This suggests in turn that there are still too few Internet education materials targeting seniors. We take a first step towards developing interactive security information materials for seniors by determining which media they prefer and can easily comprehend. We studied the reception of two media, text and audio, as they communicated information about email-based phishing attacks. Our preliminary study of 34 seniors shows that the participants personally preferred the text over the audio. However, the comprehension score was not significantly different for participants who read the phishing training text script as compared to the participants who listened to the phishing training audio script.

\end{abstract}



\section{Introduction}
The Pew Research Center reported in 2014 on the ever-increasing number of senior citizens moving to use the Internet. Safety and security are primary objectives for the growing number of seniors using the Internet: currently, 59\% of seniors over 65 use the Internet \cite{smith2014older} and, given that seniors are the fastest growing demographic, this cohort should increase annually.  But online seniors are vulnerable seniors. Researchers identify user concerns about areas such as website disclosure about purchasing history, browsing patterns or personally identifiable information, and scams, phishing and malware, and financial scams \cite{deevy2012scams}. Seniors are apparently highly vulnerable as many have low computer literacy, low awareness of Internet pitfalls, and even less knowledge about where to find information about Internet security. 

Existing HCI research mostly focuses on younger adults and university students, and rarely includes participants aged 60 and above \cite{dickinson2007methods}. Many factors contribute towards this. First, older adults are an extremely diverse group. They have significantly different lifestyle characteristics from the younger adults since most of them live far from universities. In addition, aging causes sensory changes such as visual and auditory perception, and cognitive changes such as working memory. Thus, they are more likely to forget instructions and take longer to reach a level of proficiency. Various mobility issues and illnesses may exist in older people that make it difficult for them to participate in research studies. Secondly, due to these inherent characteristics of older adults, important considerations need to be incorporated in the experimental design in order to get high quality results from them. For example, the use of flexible timing, cognitive testing, and instructions for getting to the research study venue all need to be built into the design. Thirdly, appropriate recruitment methods need to be employed in order to ensure access to a useful sample of older people. 

In order to develop interactive materials related to Internet security for seniors, it is important to first determine the seniors' media preferences. However, to the best of our knowledge, there is no existing study that explores seniors' media preference for receiving training involving computer and Internet security information, and their comprehension of each medium.

We conducted a pilot study on seniors' media preference for receiving computer and Internet security information. We focused on two media types: print text and audio, and designed scripts to communicate information about email-based phishing, and tips they can use to protect themselves against such scams. We chose email-based phishing, since it is one of the main tools used for financial fraud. Accordingly, our study focused on answering the following research questions:
\\1. Which of the two media types results in better comprehension of email-based phishing?
\\2. Which of the two media types do seniors prefer?

\section{Study Design} \label{studydesign}
Our study, approved by the UNCC Institutional Review Board\footnote{Approved IRB Protocol\#14-04-11}, followed a between-subjects design. The two treatment conditions were:
\\ \textbf{Treatment 1 - Text:} The participants read the text script on email-based phishing.
\\ \textbf{Treatment 2 - Audio:} The participants listened to an audio script on email-based phishing. The audio script was a screencast/voice-over the text script and had essentially the same contents as the text script. 

After reading/listening to the script, the participants completed a phishing comprehension survey. After completing the survey, the participants were required to listen to/read the other media script for the purpose of providing their media preference and rating of the provided scripts. The number of male and female participants in each treatment group was controlled for. 

\section{Surveys}

\subsection{Phishing comprehension}
The phishing comprehension survey consisted of a total of 10 questions. The first five questions asked the participants to label each shown email message as a legitimate or a phishing email. The next five questions were designed to test other information provided in the text and audio scripts.
Based on the responses, a score was computed out of 10. 

\subsection{Media preference and rating}
The participants answered the following questions:
\\\textbf{Q1.} Rate the audio message (Likert scale: 1 - 5)
\\\textbf{Q2.} Rate the text message (Likert scale: 1 - 5)

\subsection{Demographics}
The demographics survey comprised of 10 questions to analyze the characteristics of our participant pool.
\section{Participants}
Initially, we planned on recruiting our participants from senior centers situated off-campus, by posting flyers on their websites \cite{charlotte-mecklenburg, laurels}. 
10 seniors contacted us and showed interest in participation. However, only two of them actually participated in the study. 

The following factors impeded our recruitment process:
\\1. Reposting the flyers did not increase the number of seniors who responded to our flyers.
\\2. Most of the seniors who showed an initial interest in the study could not participate later on due to health issues or personal commitments.
\\3. A few senior centers did not give us direct permission to visit the seniors in the computer class at the senior center and interview them. Therefore, they kept us waiting for response from their senior managers.

Due to the difficulties we experienced while recruiting participants from the senior centers, we recruited our participants from Amazon Mechanical Turk (a crowdsourcing marketplace). 
We set up our study as a Human Intelligent Task (HIT), which included the tasks described in Section \ref{studydesign}. Due to the fact that a small percentage of older adults uses Amazon Mechanical Turk, we reduced the minimum eligible from 65 to 55. To ensure that only the people aged 55 and above attempt the survey, we set up our demographic survey as an eligibility screening survey. Only the participants who selected 55 and older as their age group in the demographic survey were asked to proceed with the HIT. To better control the quality of the recruited participants, we mandated that each worker has a 90\% HIT approval rating, or better. The HIT took approximately 30-40 minutes to complete, for which each worker was paid a fee of \$1. A total of 34 participants (17 per group) successfully completed the pilot study. 
\section{Results}
\subsection{Phishing Comprehension (Text script vs Audio script)}
An overall comprehension score was calculated based on the number of phishing related questions that were answered correctly (out of 10). We conducted the Wilcoxon-Mann Whitney test on the phishing comprehension scores of the two treatment groups. The test showed no significant difference between the phishing comprehension score of the text script ($\mu$=7.3, $\sigma$=1.57) and the audio script ($\mu$=8.29, $\sigma$=1.96) with p= 0.07.

\subsection{Media Preference}
Chi-squared test was conducted between the media preference for the two participant groups. The test showed that both groups have similar media preference since the p value was greater than 0.05. Both groups preferred the text media over the audio--70.5\% participants in the first group, and 64.7\% participants in the second group.
We also conducted a Wilcoxon signed-rank test between the overall Likert scale ratings of both messages. The test showed no significant differences in the ratings for text ($\mu$=3.9, $\sigma$=1.19) and audio ($\mu$=3.7, $\sigma$=1.13) message with p=0.44. 
\section{Related Work}
Garg et al. \cite{garg2012risk} studied the effectiveness of narrative-driven videos vs text for communicating phishing and malware email-based online risk to older adults. Their pilot study on 12 participants showed that video helped the participants in verbalizing the risk of responding or not responding to the emails. However, both the video and text made the participants rate the risk of responding to emails higher than that of not responding to them.
\section{Conclusion and Implications}
Our pilot study on seniors' media preferences for instructional material suggests that Dickinson's findings regarding seniors' recruitment still hold true. Dickinson states that issues such as illness and family responsibilities make it hard to recruit and schedule sessions with the seniors. Most of the seniors who initially showed interest in our study, could not participate later for similar reasons. We hoped to get a large turnout from the senior centers. In future, we plan to work with local charities and offer free computer classes (in exchange for participation) as suggested by Dickinson. 

People use the Internet for interpersonal reasons, to pass time, seek information and be entertained; MAIN, a new model for technology affordances, suggests that the visual component of multimedia surpasses text in terms of informational content delivery; however, a visual can also be seen as a distractor \cite{sundar2013uses}. New models of technology usability and gratification often lack focus on the newly emerging audience of seniors. We originally hypothesized that the empirical findings of our pilot would be able to support theoretical exploration of media acceptance by seniors and thereby enable us to tailor expanded interactive materials about security to their preferences. However, the barriers we encountered in recruiting seniors suggest that we need to develop alternative ways to recruit participants before seeking funding. Accordingly, we have revised our recruiting to include families with members placed into adult day care and caregivers for homebound cognitively impaired seniors, to develop materials and identify preferences for our VA-funded project, StoryCall, and are considering targeting churches which, particularly for minorities, often present health and computer education for its parishioners.

\balance{} 

\bibliographystyle{SIGCHI-Reference-Format}
\bibliography{sample}
\section{Appendix}

%
\textbf{Participant Demographics}
\begin{table}[!htp]
	\centering
	\renewcommand{\arraystretch}{0.75}
	\begin{tabular}{|l|l|c|}
	    \hline
		\multicolumn{2}{|c|}{\textit{Demographics} }& \textit{No. of}\\
		&&\textit{Participants}\\
		\hline		
		&&\\
		Education & High School &3\\
		&2 years of college & 8\\
		& 4 years of college & 11\\
		&$>$ 4 years of college&12\\
		&&\\
		\hline
		&&\\
		Social media use & Facebook  & 24\\
		& Twitter & 5\\
		&LinkedIn&1\\
		&None&4\\
		&&\\	
		\hline
		&&\\
		Are you concerned   & Yes  & 29\\
		about security and  & No & 5\\
		privacy when using &&\\
		the Internet&&\\	
		&&\\
		\hline
		&&\\
		Which device are you &Laptop&19\\
		using to complete this &PC&14\\
		survey&Smartphone&1\\
		&&\\
		\hline	
		&&\\
		How long have you been &Less than 5 years&6\\
		using the Internet &More than 5 years&28\\
		&&\\
		\hline	
		&&\\
		Number of times you &0&12\\
		have been a victim of &1&9\\
		Internet attack/scam	&More than 1&13\\
		&&\\
		\hline		
	\end{tabular}
	\caption{Participant Demographics}~\label{tab:demo_studies}
\end{table}

\textbf{Scripts for Education on Email-Based Phishing}
\\Text script link: https://goo.gl/noeybq
\\Audio script link: https://goo.gl/6oj86n

\textbf{Surveys}
\\Phishing Comprehension: https://goo.gl/Y4Tdc3
\\Media Preference: https://goo.gl/GfSfkX

\begin{table}[!htp]
	\centering
	\renewcommand{\arraystretch}{0.75}
	\begin{tabular}{|l|l|c|c|}
	    \hline
		\multicolumn{2}{|c|}{\textit{Demographics} }& \textit{No. of participants} & \textit{No. of participants}\\

		& &\textit{(Text Script Group )} & \textit{(Audio Script Group )}\\
		\hline		
		&&&\\
		Gender & Female & 11&8\\
		& Male & 6&9\\
		&&&\\
		\hline
	\end{tabular}
	\caption{Participant distribution within groups}~\label{tab:demo_studies}
\end{table}

\end{document}